\documentclass[aps,prd,twocolumn,showpacs,preprintnumbers,amsmath,amssymb]{revtex4-1}
\usepackage{graphicx}
\usepackage[usenames]{color}

\begin{document}
\preprint{NSF-KITP-12-027}

\title{Remarks about Dyson's instability in the large-$N$ limit}
\author{Y. Meurice}
\affiliation{KITP\\University of California\\Santa Barbara, CA 93106-4030, USA\\and\\Department of Physics and Astronomy\\ The University of Iowa\\
Iowa City, Iowa 52242, USA }

\def\lt{\lambda ^t}

\begin{abstract}
There are known examples of perturbative expansions in the 't Hooft coupling $\lt$ with a finite radius of convergence. 
This seems to contradict Dyson's argument suggesting that the instability at negative coupling implies a zero radius of convergence.
Using the example of the linear sigma model in three dimensions, we discuss to which extent the two points of view are compatible. We show that a saddle point persists 
for negative values of $\lt$ until a critical value -$|\lt_c|$ is reached. 
A numerical study of the perturbative series for the renormalized mass confirms an expected singularity of the form $(\lt +|\lt_c|)^{1/2}$.
However, for $-|\lt_c|< \lt <0 $, the effective potential does not exist if $\phi^2 >\phi^2_{max}(\lt)$ and not at all if $\lt<-|\lt_c|$. We show that 
$\phi^2_{max}(\lt)\propto 1/|\lt |$ for small negative $\lt$.  
The finite radius of convergence can be justified if the effective theory is defined with a large field cutoff $\phi^2_{max}(\lt)$ which  
provides a quantitative measure of the departure from the original model considered.
\end{abstract}
\maketitle

\section{Introduction}
The large-$N$ limit plays an important role in our understanding of gauge and spin models. It is also crucial for  
the justification of the so-called AdS/CFT correspondence. In this context, a very interesting connection between instabilities of the 
de Sitter space on the gravitational side and Dyson's instability on the gauge theory side has been suggested \cite{AM2008199}. 
In a more general context, the compatibility of Dyson's instability with perturbative series having a finite radius of convergence in the large-$N$ limit 
often leads to animated 
discussions. To the best of our knowledge, this has never been 
discussed systematically in the literature. The goal of this article is to provide such discussion together with an example where explicit  calculations are possible. 

The following observations lead to an apparent paradox. 
There are known examples of perturbative series in the 't Hooft coupling $\lt$ with a finite radius of convergence \cite{PhysRevD.7.2911,Brezin:1977sv}.
The fact that the series converges for some negative values of $\lt$ is in apparent conflict with Dyson's argument \cite{Dyson52} which generically 
suggests a zero radius of convergence due to the instability at negative coupling. This apparent conflict is called ``Dyson's large-$N$ paradox'' 
hereafter. 

There are two possible points of view. On one hand, one could say that the large-$N$ limit provides a regularization of the divergence of 
perturbative series. On the other hand, this regularization can only be obtained at the price of a ``mutilation'' \cite{Brezin:1977sv} of the 
large field contributions in the functional integral. In the following, we provide a model calculation where the two points of views 
can be discussed quantitatively. 

We consider the {\it linear} $\sigma$-model with a $\lambda (\phi^2)^2$ interaction in three dimension (3D) and a sharp momentum cutoff \cite{PhysRevD.10.2491,Arefeva:1977bt,Arefeva:1979bd,david84,david85}. For the linear $\sigma$ model, the fields can take arbitrarily large values and, at least at large but finite $N$, a general argument \cite{convpert,optim03} can be used to infer that the radius of convergence of the perturbative series should be zero. For the nonlinear version, 
the fields belong to a unit sphere of dimension $N-1$ and the situation is more complicated, as discussed in Ref. \cite{Meurice:2009bq} for a lattice regularization. Only the linear $\sigma$-model is discussed hereafter. 

The paper is organized as follows. 
In Sec. \ref{sec:paradox}, we state Dyson's large-$N$ paradox more precisely and review the scant literature on the subject. 
The large-$N$ limit of the linear $\sigma$-model is reviewed in Sec. \ref{sec:model}. A perturbative series for the renormalized mass in the symmetric phase is constructed  in Sec. \ref{sec:pert}. 
A numerical analysis of the ratios of successive coefficients provides strong evidence for a finite radius of convergence $|\lt_c|$ and a square-root singularity. 
This can be explained  by the disappearance of saddle points for $\lt<-|\lt_c|$. 
The effective potential is calculated in Sec. \ref{sec:effpot}. It is shown that for for $-|\lt_c|< \lt <0 $, the definition of the effective theory requires a large field cutoff denoted $\phi^2_{max}(\lt)$ which 
provides a measure of the mutilation mentioned above and is described as a function of $\lt$ in Sec. \ref{sec:resol}. 
The implications of these findings and possible improvements are discussed in the conclusions.

\section{Statement of the paradox}
\label{sec:paradox}
In this section, we state the nature of Dyson's large-$N$ paradox. 
For the general discussion, we use the generic notation $\lt$ for the 't Hooft coupling which is $\lambda N$ (or $g^2 N$) for scalar (for gauge, respectively) coupling and is kept constant in the large-$N$ limit. For scalar models, it is  well-known  \cite{PhysRevD.7.2911,PhysRevA.7.2172,Brezin:1977sv}, that at leading order in the large-$N$ limit, the 1-Particle Irreducible functions can be expressed 
as sums of bubble diagrams. This results 
is series with a finite radius of convergence in $\lt$ and a singularity on the negative real axis.  Explicit determination of such singularities can be found, for instance in Ref. \cite{Brezin:1977sv}. 

It was noticed by Wilson \cite{PhysRevD.7.2911} that when the quartic coupling is negative ``one expects the theory to have no ground state $ [ \dots ] $. This difficulty does not seem to show up, however, when we sum only bubble graphs." This sentence could suggest that the use of the large-$N$ limit provides a regularization of the divergence of the 
perturbative series, however, the use of ``seem'' indicates that this is not the end of the story. 

On the other hand, the fact that we obtain a converging series for values of the coupling where the theory does not make sense indicates some limitation of the approximation. Some researchers may be acquainted with this fact, however the only mention we found in the literature is in the introduction of Ref.  \cite{Brezin:1977sv}, where it is mentioned that the analyticity near the origin "reveals that the large field 
region of the Feynman path integral has been drastically mutilated". In Secs. \ref{sec:effpot} and \ref{sec:resol}, we will provide a quantitative measure of this mutilation for the specific model discussed in the next section.

\section{The model and its saddle points for $\lt>0$}
\label{sec:model}
The two points of view presented in the previous section will be discussed for the 3D linear $O(N)$ $\sigma$-model with a $\lambda (\phi^2)^2$ interaction and a sharp momentum cutoff. 
The partition function reads:
\begin{equation}\nonumber
Z=\int \mathcal{D}\phi {\rm e}^{-\int d^3x[(1/2)(\partial \phi)^2+(1/2)m_B^2\phi^2+\lambda   (\phi^2)^2-\sqrt{N}\vec{J}\vec{\phi}]}\ .
\end{equation}
Most vector products are implicit ($\phi^2\equiv \vec{\phi}.\vec{\phi}$ etc ...), $ \mathcal{D}\phi$ refers to path-integration of 
$\vec\phi$ over $\mathbb{R}^N$ rather than over the unit sphere (as in the nonlinear version) and the source $\vec{J}$ is  $x$-independent. 
We now follow closely the notations of \cite{david84}. 
We introduce auxiliary fields $M^2(x)$ enforcing the condition $\phi^2(x)=NX(x)$ and then integrate over $\phi$. Using cutoff units, 
introducing the 't Hooft coupling $\lt =\lambda N$ and dropping the 
non-zero modes (which is justified in the large-$N$ limit), the action $\mathcal{A}$ per volume and number of fields reads:
\begin{eqnarray}
\label{eq:mxaction}
\mathcal{A}=&(1/2)&\int_{|k|\leq 1}d^3k\ {\rm ln}(k^2+M^2)\\
&+&(1/2)(m_B^2-M^2)X +\lt X^2-(1/2)J^2/M^2 \nonumber \ .
\end{eqnarray}
The saddle point equations obtained by varying $X$ and $M^2$ read:
\begin{eqnarray}
\label{eq:spm}
M^2&=&m^2_B+4\lt X, \\
\label{eq:spx}
X&=&F(M^2) +J^2/(M^2)^2, 
\end{eqnarray}
with the cutoff one-loop function 
\begin{equation}
F(M^2) \equiv \int_{|k|\leq 1}\frac{d^3k}{(2\pi)^D}\frac{1}{k^2+M^2}.
\end{equation}

\begin{figure}
\includegraphics[width=2.3in,angle=0]{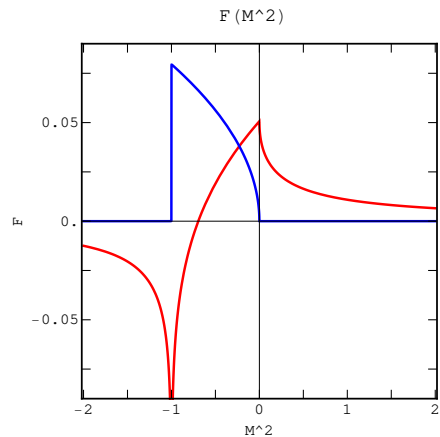}
\caption{Real part of $F(M^2)$ (red online) and imaginary part of $F(M^2-i\epsilon)$ (non-zero for $-1<M^2<0$, blue online). 
\label{fig:F}}
\end{figure}

When $M^2<0$, it is possible to analytically continue $F(M^2)$. When $-1<M^2<0$, $F(M^2)$ picks up an imaginary part. Its sign depends on the way the  path goes around the pole. 
The situation is displayed in Fig.  \ref{fig:F}, where we have used the prescription $M^2-i\epsilon$. 

\begin{figure}
\includegraphics[width=2.3in,angle=0]{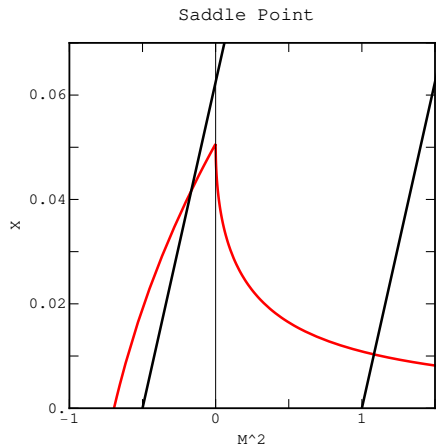}
\caption{Real part of $F(M^2)$ (red online) and the lines of $X(M^2)$ from Eq. (\ref{eq:line}) with $m_B^2=-0.5$ and $\lt$=2 (left) and $m_B^2=1$ and $\lt$=2 (right). 
\label{fig:SPSB}}
\end{figure}

The solutions of the saddle point equations (\ref{eq:spm}) and (\ref{eq:spx}), and  the phase diagram for $\lt>0$ can be obtained with the help of Fig. \ref{fig:SPSB}. When $m_B^2>0$ and $\lt>0$, the line 
\begin{equation}
\label{eq:line}
X(M^2)=(M^2-m_B^2)/(4\lt)\  ,
\end{equation}
coming from Eq. (\ref{eq:spm}) crosses $F(M^2)$ at positive $M^2$ and there is exactly one solution of Eq. (\ref{eq:spx}) when 
the source goes to zero. This situation corresponds to the symmetric phase.  

When $m_B^2<0$ and $\lt>0$, the  line 
$X(M^2)$ does not  cross $F(M^2)$ at positive $M^2$. We can nevertheless  obtain a solution of Eq. (\ref{eq:spx}) with $M^2
\rightarrow 0$ and $J^2\rightarrow 0$ with $J^2/(M^2)^2=\phi^2$ kept at the constant value $X(0)-F(0)$. This gap at $M^2=0$ is illustrated on  Fig. \ref{fig:SPSB}. This corresponds to the broken symmetry phase. The solution for $\phi^2=0$ with $M^2<0$ is briefly discussed in Sec. \ref{sec:effpot}.
\section{Perturbative series with a finite radius of convergence}
\label{sec:pert}

\begin{figure}
\includegraphics[width=2.3in,angle=0]{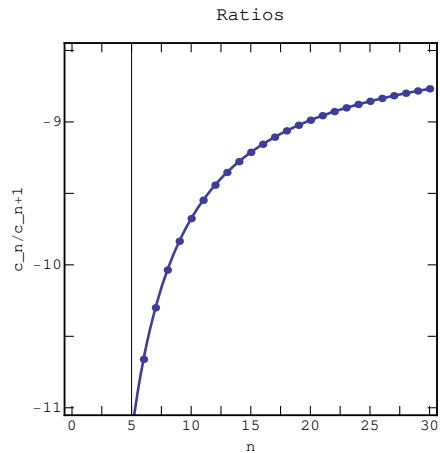}
\caption{
\label{fig:rat} 
Ratios $c_{n}/c_{n+1}$ and the fit described in the text.}
\end{figure}

In this section, we construct explicitly a perturbative series in $\lt$ for the renormalized mass in the symmetric phase. 
From the action Eq. (\ref{eq:mxaction}), we see that 
in the symmetric phase, the saddle point solution for $M^2$ in the limit of zero source, is the inverse of the zero-momentum 2-point function and we call this value of $M^2$ the renormalized mass $M^2_R$. 
Using Eqs. (\ref{eq:spm}) and (\ref{eq:spx}) in the zero source limit, we obtain the ``self-consistent" integral equation for $M^2$:
\begin{equation}
\label{eq:selfcons} 
M^2_R=m_B^2+4\lt F(M^2_R)\  .
\end{equation}
This equation can be solved perturbatively by plugging the expansion 
\begin{equation}
M^2_R=m_B^2+\sum_{n=1}^{\infty}c_n(m^2 _B)(\lt ) ^n \  ,
\end{equation}
in Eq. (\ref{eq:selfcons}). One finds $c_1(m^2 _B)=4F(m_B^2)$ and so on. 
The series has been calculated numerically up to order 30 for $m^2_B=1$. The ratio of successive coefficients is showed in Fig. \ref{fig:rat}. The last 25 ratios have been fitted as follows:
\begin{equation}
c_n/c_{n+1}\simeq -8.34356-12.3454/n  -8.73165/n^2 \ .
\end{equation}
We can compare this behavior with the corresponding ratios for the series of $(x-x_c)^\alpha$ which read 
\begin{equation}
c_n/c_{n+1}\simeq x_c(1+(1+\alpha)/n+\mathcal{O}(1/n^2)) \ .
\end{equation}
We conclude that the series for $M_R^2(\lt)$  has a singularity at $\lt \simeq -8.344$ and a power behavior 
$\alpha\simeq 0.493$.  These two approximate numbers have a simple interpretation.

\begin{figure}
\includegraphics[width=2.3in,angle=0]{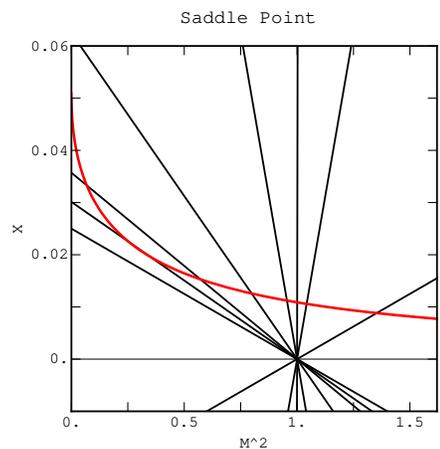}
\caption{Real part of $F(M^2)$ (red online) and the lines of $X(M^2)$ from Eq. (\ref{eq:spx}) with $m_B^2=1$ and $\lt$=
10, 1, 0.01, -1, -4, -7, -8.3 and  -10 (couterclockwise). 
\label{fig:SP}}
\end{figure}

The singularity near -8.344 can be explained in terms of the saddle points properties when $\lt <0$. The solutions of the 
zero source, symmetric phase, saddle point equations can be read from Fig. \ref{fig:SP} where we have displayed the lines $X(M^2)$ for $m_B^2=1$ and a certain number of values of $\lt$ ranging from 10 to -10. For positive 
$\lt$ there is only one intersection between the lines and $F(M^2)$. As $\lt$ becomes negative but not too negative, this property persists. When $\lt$ is decreased to approximately -4.935, a second solution appears at lower (but positive) $M^2$. 
As $\lt$ further decreases, the two solutions get closer and finally coalesce for $\lt = -8.3419 ...$ and disappear for more negative values. This provides strong numerical evidence that the singularity of the perturbative series is due to the absence of saddle point beyond this critical value. 

The nature of the singularity can be further understood using the strong coupling limit of the self-consistent Eq. (\ref{eq:selfcons}). By inspection, it is clear that when $\lt$ becomes large so does $M^2$. In this limit, we can neglect $m_B^2$ and 
the $k^2$ at the denominator of $F(M^2)$. This implies 
\begin{equation}
M^2_R\simeq \sqrt{2\lt/(3\pi^2)} .
\end{equation}
This square root behavior is in very good agreement with the estimate of the power of the singularity $\alpha \simeq 0.493$. 

\section{The effective potential}
\label{sec:effpot}

The effective potential $V_{eff}(\vec{\phi})$ can be constructed by a Legendre transformation \cite{PhysRevD.10.2491,Arefeva:1977bt,david84}. The source can be eliminated through the relation 
$\vec{J}=M^2 \vec{\phi}$
and we obtain
\begin{eqnarray}
V_{eff}(\vec{\phi})=(&1/2&) (m_B^2-M^2)X+\lt X^2\\ \nonumber &+&(1/2)M^2\phi^2+\int_{|k|\leq 1}d^3k \ {\rm ln}(k^2+M^2)\ .
\end{eqnarray}
The saddle point equations are Eq. (\ref{eq:spm}) and 
\begin{equation}
\label{eq:spXphi}
X=F(M^2) +\phi^2 \  .
\end{equation}

The values of $V_{eff}(\vec{\phi})$ can be constructed by using $M^2$ as a parameter. For fixed values of $m_B^2$ and $\lt$, choosing a value of $M^2$ determines 
$X(M^2)$ using Eq. (\ref{eq:line}) as before.  Plugging this expression in Eq. (\ref{eq:spXphi}), we obtain $\phi^2(M^2)$ which is positive provided that $X(M^2)>F(M^2)$. For $m^2_B>0$ and $\lt >0$ (symmetric phase), this implies $M^2>M^2_R$ as discussed in Sec. \ref{sec:model} and shown in Fig. \ref{fig:SPSB}. The important point is that in this case all positive values of $\phi^2$ are allowed. On the other hand, for $m^2_B<0$ and $\lt >0$ (broken symmetry  phase), $M^2$ can take any positive value but $\phi^2\geq X(0)-F(0)>0$ as illustrated in Fig. \ref{fig:SPSB}. Smaller values of 
$\phi^2$ could in principle be reached by using negative values of $M^2$, but then an imaginary part appears signaling an instability \cite{PhysRevD.10.2491}. This can be understood with a simple double-well potential example $(\phi^2-v^2)^2$: if a source is introduced the absolute 
minimum is always for $\phi^2>v^2$ and the extrema for $\phi^2<v^2$ correspond to subdominant saddles. For this reason we only consider the case $M^2\geq 0$ in the following.

\begin{figure}
\includegraphics[width=2.3in,angle=0]{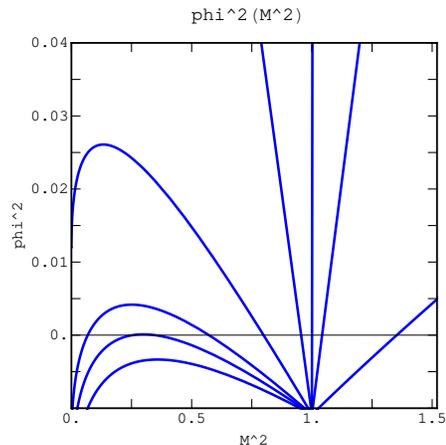}
\caption{$\phi^2(M^2)$ for  $\lt$=
10, 1, 0.01, -1, -4, -7, -8.3 and  -10 (conterclockwise). The correct range of $M^2$ is selected by requiring the positivity of $\phi^2$. 
\label{fig:phiofm}}
\end{figure}

\begin{figure}
\includegraphics[width=2.3in,angle=0]{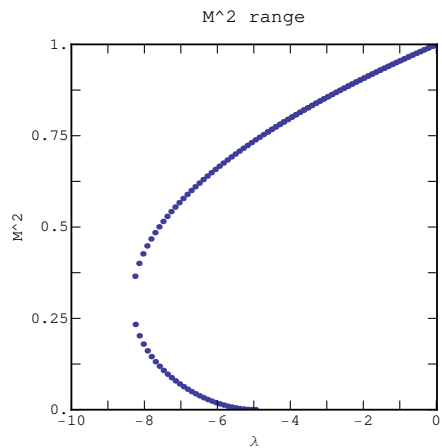}
\caption{Boundary of the values of $M^2$ insuring $\phi^2>0$ as a function of $\lt$. The lower right corner is inside the boundary.
\label{fig:msbound}}
\end{figure}

We now restrict the discussion to $m^2_B>0$ and vary $\lt$ from positive to negative values. We use the same numerical values as in Sec. \ref{sec:pert}. The curves $\phi^2(M^2)$ are shown in Fig. \ref{fig:phiofm}. The three curves on the right correspond to $\lt >0$ and illustrate 
that in this case, $\phi^2$ can take arbitrarily large positive values. The five curves on the left correspond to $\lt<0$ and have a maximal value 
of $\phi^2$. For $\lt =-1$ this maximum is reached outside of the figure. For $\lt$ not too negative, $\phi^2$ remains positive when $M^2$ goes to zero. If $\lt$ is decreased below the approximate value -4.935, $\phi^2$ goes back to zero before $M^2$ is zero. The correct range of $M^2$ is then 
selected by requiring the positivity of $\phi^2$ as illustrated in Fig. \ref{fig:msbound}. We can now vary $M^2$ within this correct range and construct a parametric representation of $V_{eff}(\vec{\phi})$.  

\begin{figure}[t]
\vskip5pt
\includegraphics[width=2.3in,angle=0]{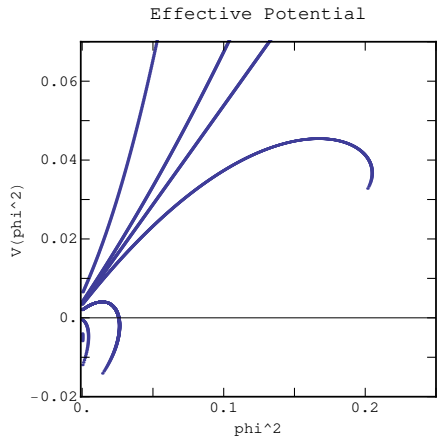}
\caption{$V_{eff}(\vec{\phi})$ for $\lt$=
10, 1, 0.01, -1, -4, -7, and, barely visible, -8.3 (clockwise). The two solutions were kept whenever present. 
\label{fig:effpot}}
\end{figure}

Fig. \ref{fig:phiofm} shows that as $M^2$ is varied, some values of $\phi^2$ will appear twice. When this is the case, we need to pick the value that has the smallest $V_{eff}(\vec{\phi})$. The effective potential corresponding to the set of values of $\lt$ discussed above, is shown in Fig. \ref{fig:effpot} where the two possible solutions, whenever present, have been kept. It should be understood that when $\phi^2$ reaches the value where there are two solutions, $V_{eff}(\vec{\phi})$ drops discontinuously to  the lowest solution. 

\section{Partial resolution of the paradox}
\label{sec:resol}

It is clear from Eq. (\ref{eq:spXphi}) and Figs. \ref{fig:SP} and \ref{fig:effpot} that when $-|\lt_c|<\lt<0$, $\phi^2$ cannot take arbitrary large values. 
In the following, we call the maximal value $\phi^2_{max} (\lt)$. For $m_B^2$ and $\lt$ fixed, $\phi^2_{max}(\lt)$ can be obtained by maximizing $\phi^2$ in Eq. (\ref{eq:spXphi}). 
This yields 
\begin{equation}
F'(M^2)=1/(4\lt).
\end{equation}

\begin{figure}
\includegraphics[width=2.3in,angle=0]{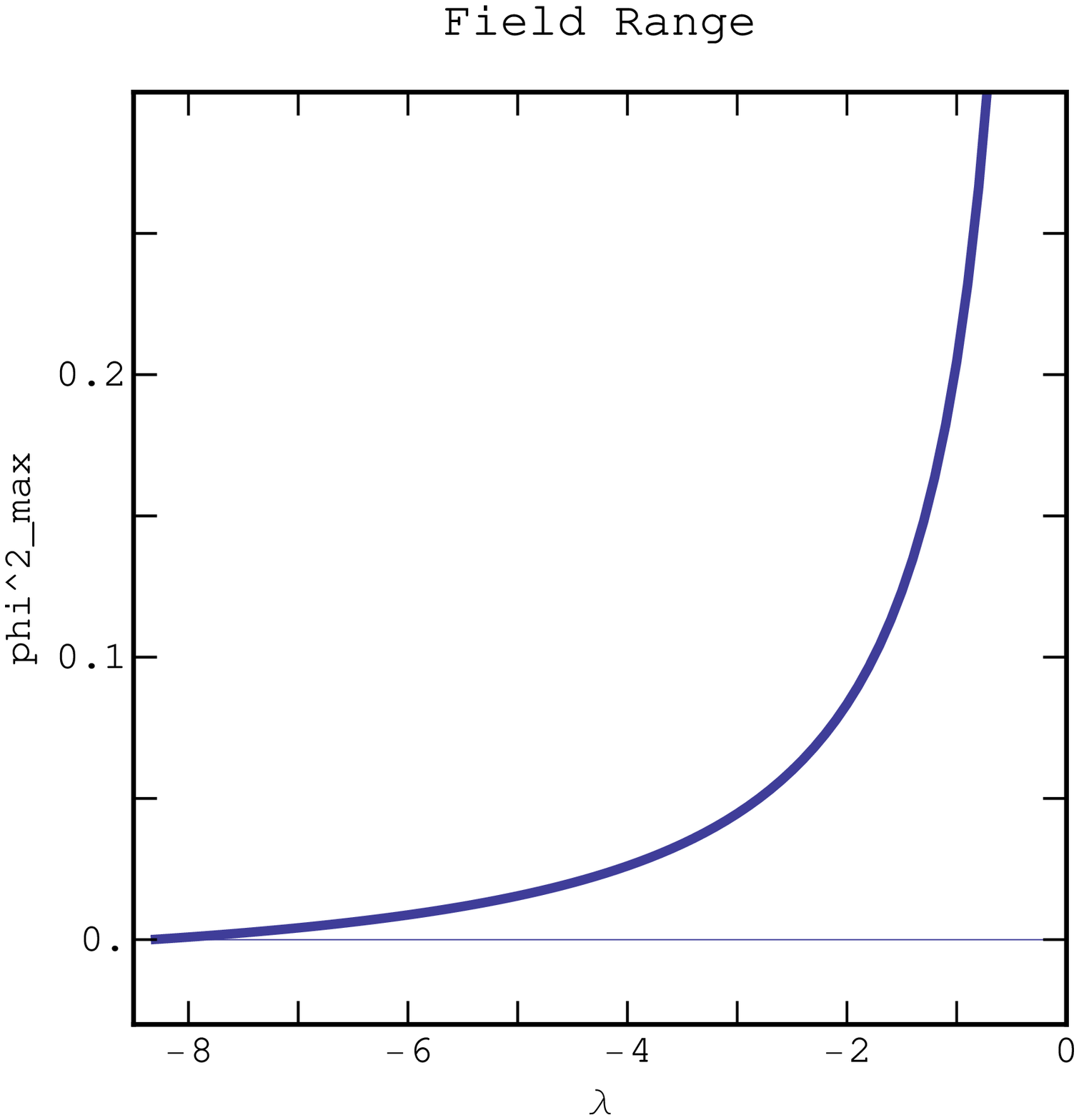}
\caption{Values of $\phi^2_{max}(\lt)$ versus $\lt$. 
\label{fig:phimax}}
\end{figure}

For negative value of $\lt$ with a small absolute value, the function $X(M^2)$ is almost vertical and the matching with the slope of $F(M^2)$ requires 
a small value of $M^2$, namely $M^2 \simeq (\lt)^2 /({4\pi^2})$. Keeping only the leading term in $X-F$, we obtain that in this limit, 
\begin{equation}\phi^2_{max}\simeq m_B^2/(4|\lt|) .
\end{equation}
On the other hand, when $\lt$ approaches $\lt_c$ from above, $\phi^2_{max}(\lt)$ goes to zero  linearly. These two behaviors are illustrated in Fig. \ref{fig:phimax}. 

An effective theory for the zero mode of the field can in principle be defined at negative $\lt$ provided that we discard the large field region 
$\phi^2> \phi^2_{max}(\lt )$. Under these circumstances, it is possible to set bounds on the perturbative coefficients \cite{convpert,optim03} of the 
partition function that guarantees that the series converges and the perturbative series have a finite radius of convergence determined 
by the zeros of the partition function. In the example considered here, the introduction of a large field cutoff represents a departure from the original model. 
Consequently, the function $\phi^2_{max}(\lt )$  can be interpreted as a measure of the mutilation of the large field contributions introduced in Sec. \ref{sec:paradox}. 

\section{Conclusions}

In summary, we have discussed quantitatively the two aspects of Dyson large-$N$ paradox for the 3D linear $O(N)$ sigma model.  
At leading order in the large-$N$ expansion, the perturbative series in $\lt$  for the renormalized mass shows very good evidence for 
a singular behavior of the form $(\lt +|\lt_c|)^{1/2}$, where $\lt_c$ is the smallest (negative) value for which a saddle point exists. 
The convergence of the series for $-|\lt_c|< \lt <0 $ can be explained by the fact that for this range of $\lt$, the effective theory can only be defined 
for $\phi^2<\phi^2_{max}(\lt )$. 

The explicit construction of the effective potential 
for $\lt$ negative but not too negative, shows that $\phi^2=0$ remains an absolute minimum of the effective potential in the range where it can be defined. This saddle point can be used in the large-$N$ limit despite the pathologies that the effective potential develops at larger $\phi^2$, namely 
a discontinuity and a finite range of definition. If $\lt$ is lowered below  $\lt_c$,  the two solutions of the saddle point equation coalesce and  disappear completely, and so does the effective potential. 
This mechanism seems to be generic and it should be possible to observe similar disappearances of large-$N$ saddle points in other models. 

More work is necessary in order to understand 
how the $1/N$ corrections affect the large field behavior of the effective potential at positive $\lt$. Effects that only appear at the $1/N$ level are particularly interesting with this respect. In the context of QCD-like theories, the breaking of the axial $U(1)$ \cite{Witten:1979vv,Veneziano:1979ec} is an example of such phenomenon. 
We also expect that a better understanding of the connection between de Sitter and anti de Sitter spaces could help us understand complex renormalization group flows \cite{Denbleyker:2010sv,PhysRevD.83.056009,Liu:2011zzh} in gauge theories and $\sigma$-models and find improved weak coupling expansions.

\begin{acknowledgments}
This article was written while attending the KITP workshop ``Novel Numerical Methods for Strongly Coupled Quantum Field Theory and Quantum Gravity".  We had many interesting conversations with many participants and especially 
L.  Pando Zayas, J. Polchinski and P. Damgaard regarding Dyson and de Sitter instabilities.  
This 
research was supported in part  by the Department of Energy
under Contract No. FG02-91ER40664 and by the National Science Foundation under Grant No. PHY11-25915. \end{acknowledgments}


\begin{thebibliography}{10}%
\makeatletter
\providecommand \@ifxundefined [1]{%
 \ifx #1\undefined \expandafter \@firstoftwo
 \else \expandafter \@secondoftwo
\fi
}%
\providecommand \@ifnum [1]{%
 \ifnum #1\expandafter \@firstoftwo
 \else \expandafter \@secondoftwo
\fi
}%
\providecommand \enquote [1]{``#1''}%
\providecommand \bibnamefont  [1]{#1}%
\providecommand \bibfnamefont [1]{#1}%
\providecommand \citenamefont [1]{#1}%
\providecommand\href[0]{\@sanitize\@href}%
\providecommand\@href[1]{\endgroup\@@startlink{#1}\endgroup\@@href}%
\providecommand\@@href[1]{#1\@@endlink}%
\providecommand \@sanitize [0]{\begingroup\catcode`\&12\catcode`\#12\relax}%
\@ifxundefined \pdfoutput {\@firstoftwo}{%
 \@ifnum{\z@=\pdfoutput}{\@firstoftwo}{\@secondoftwo}%
}{%
 \providecommand\@@startlink[1]{\leavevmode}%
 \providecommand\@@endlink[0]{}%
}{%
 \providecommand\@@startlink[1]{%
  \leavevmode
  \pdfstartlink
   attr{/Border[0 0 1 ]/H/I/C[0 1 1]}%
   user{/Subtype/Link/A<</Type/Action/S/URI/URI(#1)>>}%
  \relax
 }%
 \providecommand\@@endlink[0]{\pdfendlink}%
}%
\providecommand \url  [0]{\begingroup\@sanitize \@url }%
\providecommand \@url [1]{\endgroup\@href {#1}{\urlprefix}}%
\providecommand \urlprefix [0]{URL }%
\providecommand \Eprint[0]{\href }%
\@ifxundefined \urlstyle {%
  \providecommand \doi [1]{doi:\discretionary{}{}{}#1}%
}{%
  \providecommand \doi [0]{doi:\discretionary{}{}{}\begingroup
  \urlstyle{rm}\Url }%
}%
\providecommand \doibase [0]{http://dx.doi.org/}%
\providecommand \Doi[1]{\href{\doibase#1}}%
\providecommand \bibAnnote [3]{%
  \BibitemShut{#1}%
  \begin{quotation}\noindent
    \textsc{Key:}\ #2\\\textsc{Annotation:}\ #3%
  \end{quotation}%
}%
\providecommand \bibAnnoteFile [2]{%
  \IfFileExists{#2}{\bibAnnote {#1} {#2} {\input{#2}}}{}%
}%
\providecommand \typeout [0]{\immediate \write \m@ne }%
\providecommand \selectlanguage [0]{\@gobble}%
\providecommand \bibinfo [0]{\@secondoftwo}%
\providecommand \bibfield [0]{\@secondoftwo}%
\providecommand \translation [1]{[#1]}%
\providecommand \BibitemOpen[0]{}%
\providecommand \bibitemStop [0]{}%
\providecommand \bibitemNoStop [0]{.\EOS\space}%
\providecommand \EOS [0]{\spacefactor3000\relax}%
\providecommand \BibitemShut [1]{\csname bibitem#1\endcsname}%
\bibitem{AM2008199}%
  \BibitemOpen
  \bibfield{author}{%
  \bibinfo {author}
  {\bibnamefont{A. M. Polyakov}},\ }%
  \bibfield{journal}{%
  {\bibinfo {journal} {Nuclear Physics
  B}}\ }%
  \textbf{\bibinfo {volume} {797}},\ \bibinfo {pages} {199 } (\bibinfo {year}
  {2008})
  \bibAnnoteFile{NoStop}{AM2008199}%
\bibitem{PhysRevD.7.2911}%
  \BibitemOpen
  \bibfield{author}{%
  \bibinfo {author} {\bibfnamefont{K.~G.}\ \bibnamefont{Wilson}},\ }%
  \bibfield{journal}{%
  \Doi{10.1103/PhysRevD.7.2911}{\bibinfo {journal} {Phys. Rev. D}}\ }%
  \textbf{\bibinfo {volume} {7}},\ \bibinfo {pages} {2911} ( \bibinfo {year} {1973})
  \bibAnnoteFile{NoStop}{PhysRevD.7.2911}%
\bibitem{Brezin:1977sv}%
  \BibitemOpen
  \bibfield{author}{%
  \bibinfo {author} {\bibfnamefont{E.}~\bibnamefont{Brezin}}, \bibinfo {author}
  {\bibfnamefont{C.}~\bibnamefont{Itzykson}}, \bibinfo {author}
  {\bibfnamefont{G.}~\bibnamefont{Parisi}},\ and\ \bibinfo {author}
  {\bibfnamefont{J.~B.}\ \bibnamefont{Zuber}},\ }%
  \bibfield{journal}{%
  \Doi{10.1007/BF01614153}{\bibinfo {journal} {Commun. Math. Phys.}}\ }%
  \textbf{\bibinfo {volume} {59}},\ \bibinfo {pages} {35} (\bibinfo {year}
  {1978})%
  \bibAnnoteFile{NoStop}{Brezin:1977sv}%
\bibitem{Dyson52}%
  \BibitemOpen
  \bibfield{author}{%
  \bibinfo {author} {\bibfnamefont{F.}~\bibnamefont{Dyson}},\ }%
  \bibfield{journal}{%
  \bibinfo {journal} {Phys. Rev.}\ }%
  \textbf{\bibinfo {volume} {85}},\ \bibinfo {pages} {631} (\bibinfo {year}
  {1952})%
  \bibAnnoteFile{NoStop}{Dyson52}%
\bibitem{PhysRevD.10.2491}%
  \BibitemOpen
  \bibfield{author}{%
  \bibinfo {author} {\bibfnamefont{S.}~\bibnamefont{Coleman}}, \bibinfo
  {author} {\bibfnamefont{R.}~\bibnamefont{Jackiw}},\ and\ \bibinfo {author}
  {\bibfnamefont{H.~D.}\ \bibnamefont{Politzer}},\ }%
  \bibfield{journal}{%
  {\bibinfo {journal} {Phys. Rev. D}}\ }%
  \textbf{\bibinfo {volume} {10}},\ \bibinfo {pages} {2491} (\bibinfo {year} {1974})
  \bibAnnoteFile{NoStop}{PhysRevD.10.2491}%
\bibitem{Arefeva:1977bt}%
  \BibitemOpen
  \bibfield{author}{%
  \bibinfo {author} {\bibfnamefont{I.~Y.}\ \bibnamefont{Arefeva}},\ }%
  \bibfield{journal}{%
  \bibinfo {journal} {Teor. Mat. Fiz.}\ }%
  \textbf{\bibinfo {volume} {31}},\ \bibinfo {pages} {3} (\bibinfo {year}
  {1977})%
  \bibAnnoteFile{NoStop}{Arefeva:1977bt}%
\bibitem{Arefeva:1979bd}%
  \BibitemOpen
  \bibfield{author}{%
  \bibinfo {author} {\bibfnamefont{I.~Y.}\ \bibnamefont{Arefeva}},\ }%
  \bibfield{journal}{%
  \Doi{10.1016/0003-4916(79)90361-0}{\bibinfo {journal} {Annals Phys.}}\ }%
  \textbf{\bibinfo {volume} {117}},\ \bibinfo {pages} {393} (\bibinfo {year}
  {1979})%
  \bibAnnoteFile{NoStop}{Arefeva:1979bd}%
\bibitem{david84}%
  \BibitemOpen
  \bibfield{author}{%
  \bibinfo {author} {\bibfnamefont{F.}~\bibnamefont{David}}, \bibinfo {author}
  {\bibfnamefont{D.~A.}\ \bibnamefont{Kessler}},\ and\ \bibinfo {author}
  {\bibfnamefont{H.}~\bibnamefont{Neuberger}},\ }%
  \bibfield{journal}{%
  \bibinfo {journal} {Phys. Rev. Lett.}\ }%
  \textbf{\bibinfo {volume} {53}},\ \bibinfo {pages} {2071} (\bibinfo {year}
  {1984})%
  \bibAnnoteFile{NoStop}{david84}%
\bibitem{david85}%
  \BibitemOpen
  \bibfield{author}{%
  \bibinfo {author} {\bibfnamefont{F.}~\bibnamefont{David}}, \bibinfo {author}
  {\bibfnamefont{D.~A.}\ \bibnamefont{Kessler}},\ and\ \bibinfo {author}
  {\bibfnamefont{H.}~\bibnamefont{Neuberger}},\ }%
  \bibfield{journal}{%
  \bibinfo {journal} {Nucl. Phys.}\ }%
  \textbf{\bibinfo {volume} {B257}},\ \bibinfo {pages} {695} (\bibinfo {year}
  {1985})%
  \bibAnnoteFile{NoStop}{david85}%
\bibitem{convpert}%
  \BibitemOpen
  \bibfield{author}{%
  \bibinfo {author} {\bibfnamefont{Y.}~\bibnamefont{Meurice}},\ }%
  \bibfield{journal}{%
  \bibinfo {journal} {Phys. Rev. Lett.}\ }%
  \textbf{\bibinfo {volume} {88}},\ \bibinfo {pages} {141601} (\bibinfo {year}
  {2002}),\ \Eprint{http://arxiv.org/abs/hep-th/0103134}{hep-th/0103134}%
  \bibAnnoteFile{NoStop}{convpert}%
\bibitem{optim03}%
  \BibitemOpen
  \bibfield{author}{%
  \bibinfo {author} {\bibfnamefont{B.}~\bibnamefont{Kessler}}, \bibinfo
  {author} {\bibfnamefont{L.}~\bibnamefont{Li}},\ and\ \bibinfo {author}
  {\bibfnamefont{Y.}~\bibnamefont{Meurice}},\ }%
  \bibfield{journal}{%
  \bibinfo {journal} {Phys. Rev.}\ }%
  \textbf{\bibinfo {volume} {D69}},\ \bibinfo {pages} {045014} (\bibinfo {year}
  {2004}),\ \Eprint{http://arxiv.org/abs/hep-th/0309022}{hep-th/0309022}%
  \bibAnnoteFile{NoStop}{optim03}%
\bibitem{Meurice:2009bq}%
  \BibitemOpen
  \bibfield{author}{%
  \bibinfo {author} {\bibfnamefont{Y.}~\bibnamefont{Meurice}},\ }%
  \bibfield{journal}{%
  \bibinfo {journal} {Phys. Rev.}\ }%
  \textbf{\bibinfo {volume} {D80}},\ \bibinfo {pages} {054020} (\bibinfo {year}
  {2009}),\ \Eprint{http://arxiv.org/abs/0907.2980}{arXiv:0907.2980 [hep-lat]}%
  \bibAnnoteFile{NoStop}{Meurice:2009bq}%
\bibitem{PhysRevA.7.2172}%
  \BibitemOpen
  \bibfield{author}{%
  \bibinfo {author} {\bibfnamefont{S.-K.}\ \bibnamefont{Ma}},\ }%
  \bibfield{journal}{%
  \Doi{10.1103/PhysRevA.7.2172}{\bibinfo {journal} {Phys. Rev. A}}\ }%
  \textbf{\bibinfo {volume} {7}},\ \bibinfo {pages} {2172} (\bibinfo {year} {1973})
  \bibAnnoteFile{NoStop}{PhysRevA.7.2172}%
\bibitem{Witten:1979vv}%
  \BibitemOpen
  \bibfield{author}{%
  \bibinfo {author} {\bibfnamefont{E.}~\bibnamefont{Witten}},\ }%
  \bibfield{journal}{%
  \Doi{10.1016/0550-3213(79)90031-2}{\bibinfo {journal} {Nucl.Phys.}}\ }%
  \textbf{\bibinfo {volume} {B156}},\ \bibinfo {pages} {269} (\bibinfo {year}
  {1979})%
  \bibAnnoteFile{NoStop}{Witten:1979vv}%
\bibitem{Veneziano:1979ec}%
  \BibitemOpen
  \bibfield{author}{%
  \bibinfo {author} {\bibfnamefont{G.}~\bibnamefont{Veneziano}},\ }%
  \bibfield{journal}{%
  \Doi{10.1016/0550-3213(79)90332-8}{\bibinfo {journal} {Nucl.Phys.}}\ }%
  \textbf{\bibinfo {volume} {B159}},\ \bibinfo {pages} {213} (\bibinfo {year}
  {1979})%
  \bibAnnoteFile{NoStop}{Veneziano:1979ec}%
\bibitem{Denbleyker:2010sv}%
  \BibitemOpen
  \bibfield{author}{%
  \bibinfo {author} {\bibfnamefont{A.}~\bibnamefont{Denbleyker}}, \bibinfo
  {author} {\bibfnamefont{D.}~\bibnamefont{Du}}, \bibinfo {author}
  {\bibfnamefont{Y.}~\bibnamefont{Liu}}, \bibinfo {author}
  {\bibfnamefont{Y.}~\bibnamefont{Meurice}},\ and\ \bibinfo {author}
  {\bibfnamefont{H.}~\bibnamefont{Zou}},\ }%
  \bibfield{journal}{%
  \Doi{10.1103/PhysRevLett.104.251601}{\bibinfo {journal} {Phys. Rev. Lett.}}\
  }%
  \textbf{\bibinfo {volume} {104}},\ \bibinfo {pages} {251601} (\bibinfo {year}
  {2010}),\ \Eprint{http://arxiv.org/abs/1005.1993}{arXiv:1005.1993 [hep-lat]}%
  \bibAnnoteFile{NoStop}{Denbleyker:2010sv}%
\bibitem{PhysRevD.83.056009}%
  \BibitemOpen
  \bibfield{author}{%
  \bibinfo {author} {\bibfnamefont{Y.}~\bibnamefont{{Meurice}}}\ and\ \bibinfo
  {author} {\bibfnamefont{H.}~\bibnamefont{{Zou}}},\ }%
  \bibfield{journal}{%
  \Doi{10.1103/PhysRevD.83.056009}{\bibinfo {journal} {\prd}}\ }%
  \textbf{\bibinfo {volume} {83}},\ \bibinfo {pages} {056009} (\bibinfo {year} {2011}),\
  \Eprint{http://arxiv.org/abs/1101.1319}{arXiv:1101.1319 [hep-lat]}%
  \bibAnnoteFile{NoStop}{PhysRevD.83.056009}%
\bibitem{Liu:2011zzh}%
  \BibitemOpen
  \bibfield{author}{%
  \bibinfo {author} {\bibfnamefont{Y.}~\bibnamefont{Liu}}\ and\ \bibinfo
  {author} {\bibfnamefont{Y.}~\bibnamefont{Meurice}},\ }%
  \bibfield{journal}{%
  \Doi{10.1103/PhysRevD.83.096008}{\bibinfo {journal} {Phys. Rev. D}}\ }%
  \textbf{\bibinfo {volume} {83}},\ \bibinfo {pages} {096008} (\bibinfo {year}
  {2011}),\ \Eprint{http://arxiv.org/abs/1103.4846}{arXiv:1103.4846 [hep-lat]}%
  \bibAnnoteFile{NoStop}{Liu:2011zzh}%
\end{thebibliography}
\end{document}